\documentclass[superscriptaddress,aps,prc,showpacs,floats,floatfix,twocolumn]{revtex4-1} 
\usepackage{epsfig}
\usepackage{amsmath}
\usepackage{amsfonts}
\usepackage{color}

\begin{document}

\renewcommand{\vec}[1]{{\mbox{\boldmath $#1$}}}

\title{Fission properties for r-process nuclei}

\author{J. Erler}
\affiliation{Institut f\"ur Theoretische Physik II, Universit\"at 
  Erlangen-N\"urnberg, Staudtstrasse 7, D-91058 Erlangen, Germany}
\affiliation{Department of Physics and Astronomy, University of
  Tennessee, Knoxville, Tennessee 37996, USA} 
\affiliation{Physics Division, Oak Ridge National Laboratory, Oak
  Ridge, Tennessee 37831, USA} 

\author{K. Langanke}
\affiliation{GSI Helmholtzzentrum f\"ur Schwerionenforschung,
  Planckstr. 1, 64291 Darmstadt, Germany}
\affiliation{Technische Universit{\"a}t Darmstadt, Institut f{\"u}r
  Kernphysik, Schlossgartenstr. 9, 64289 Darmstadt, Germany}
\affiliation{Frankfurt Institute of Advanced Studies, Ruth-Moufang
  Str. 1, 60438 Frankfurt, Germany}

\author{H.P. Loens} 
\affiliation{GSI Helmholtzzentrum f\"ur Schwerionenforschung,
  Planckstr. 1, 64291 Darmstadt, Germany} 
\affiliation{Technische Universit{\"a}t Darmstadt, Institut f{\"u}r
  Kernphysik, Schlossgartenstr. 9, 64289 Darmstadt, Germany}

\author{G. Mart\'inez-Pinedo}
\affiliation{Technische Universit{\"a}t Darmstadt, Institut f{\"u}r
  Kernphysik, Schlossgartenstr. 9, 64289 Darmstadt, Germany}
\affiliation{GSI Helmholtzzentrum f\"ur Schwerionenforschung,
  Planckstr. 1, 64291 Darmstadt, Germany} 

\author{P.--G. Reinhard} 
\affiliation{Institut f\"ur Theoretische Physik II, Universit\"at
  Erlangen-N\"urnberg, Staudtstrasse 7, D-91058 Erlangen, Germany}

\date{\today}

\begin{abstract}
  We present a systematics of fission barriers and fission lifetimes
  for the whole landscape of super-heavy elements (SHE), i.e. nuclei
  with $Z\geq 100$. The fission lifetimes are also compared with the
  $\alpha$-decay half-lives. The survey is based on a self-consistent
  description in terms of the Skyrme-Hartree-Fock (SHF)
  approach. Results for various different SHF parameterizations are
  compared to explore the robustness of the predictions.  The fission
  path is computed by quadrupole constrained SHF.  The computation of
  fission lifetimes takes care of the crucial ingredients of the
  large-amplitude collective dynamics along the fission path, as
  self-consistent collective mass and proper quantum corrections.  We
  discuss the different topologies of fission landscapes which occur
  in the realm of SHE (symmetric versus asymmetric fission, regions of
  triaxial fission, bi-modal fission, and the impact of asymmetric
  ground states). The explored region is extended deep into the regime
  of very neutron-rich isotopes as they are expected to be produced in
  the astrophysical r process. 
\end{abstract}

\maketitle

\section{Introduction}

The existence of super-heavy elements (SHE) above the naturally
existing ones has attracted much attention in the past decades
\cite{Nix72aR,Hof00aER}.  The topic remains of high actual interest as
the new and heavier synthesized elements are added every year to the
list, for a few examples from the rich list see
\cite{HofS01,Oga04,Oga06,Greg06,Dvo08}. Super-heavy elements are also
produced during the
r-process~\cite{Martinez-Pinedo.Mocelj.ea:2007,Arn07,Sto07aR} and
their properties are important in order to determine the upper end of
the nucleosynthesis flow. The key question in the study of SHE is
their stability against the various decay channels as $\alpha$-decay,
$\beta$-decay, and particularly spontaneous fission. This paper aims
at a theoretical survey of fission lifetimes for SHE. It will
establish a systematics all over the landscape of SHE from the
experimentally accessible neutron poor ones to the very neutron rich
species which may occur in the r process. Fission lifetimes will also
be compared with the lifetimes for $\alpha$-decay. This survey is
based on a theoretical description at level of a self-consistent mean
field (SCMF). Such models came into practice about 40 years ago and
have been steadily developed to deliver now a reliable
description of nuclear structure and dynamics, for recent reviews
see~\cite{Rin96aR,Ben03aR,Vre05aR,Sto07aR,Erl11aR}. We use here in
particular the Skyrme-Hartree-Fock (SHF) approach which stays in the
non-relativistic domain and employs an effective energy functional
corresponding to zero-range interactions \cite{Ben03aR}.

The first theoretical estimates of fission stability were performed on
the grounds of the shell correction energy within phenomenologically
adjusted shell model potentials \cite{Bra72aR} and studies of shell
structure persist to be of high exploratory value also for
self-consistent approaches \cite{Ben99a,Kru00a}. In the realm of SHE,
one finds broad islands of shell stabilization rather than the narrow
and deep valleys as they are typically found for lighter nuclei
\cite{Ben01a}. The emergence of large regions of stable nuclei is, in
fact, favorable for the potential experimental accessibility. The next
step after estimates from the shell correction energy is to check the
systematics of fission barriers in SHE.  There exists a wealth of
information about fission in actinide nuclei \cite{Spe74aER} which
helps to probe the predictive value of the theoretical approaches,
e.g. of the SHF method \cite{Flo74a,Bue04a}. (In fact, a fission
barrier was used in the calibration of one SHF functional
\cite{Bar82a}.) However, comparison with experimental data requires to
go beyond a pure SHF description and to take into account collective
correlations (from rotation and low-energy vibration) which can modify
the fission barriers by up to 2 MeV \cite{Rei87aR,Rei76a}.  The
systematics of fission barriers in SHE is simplified by the fact that
there is only one fission barrier to be considered (as opposed to
actinides with their double-humped barrier).  It was found
\cite{Ben98a,Bue04a} that SHF provides estimates of islands of fission
stability which are qualitatively in accordance with experiments
\cite{Oga99aE,Oga99cE,Oga01aE}. The ultimate goal is, of course, to
estimate the fission lifetimes directly.  However, self-consistent
calculations of fission lifetimes are extremely demanding and thus
have come up only recently, see e.g. \cite{Ber01a,War06,Sta09a} (which 
are mostly using still approximate masses and quantum corrections
\cite{Rei87aR}) or \cite{Sch09a} for a fully self-consistent
calculation.  In this paper, we
employ the method as presented in \cite{Sch09a} for establishing the
systematics of fission lifetimes for all conceivable SHE. We will
discuss the influence of the choice of the SHF parameterization and we
will compute and compare also the lifetimes for $\alpha$ decay.

The paper is outlined as follows: Section~\ref{sec:formal} presents
the formal framework for the computation of fission lifetimes.
Section \ref{sec:results} discusses a variety of results on fission
barriers, fission lifetimes and a comparison with $\alpha$-decay
lifetimes.

\section{Formal framework}
\label{sec:formal}

\subsection{SHF approach and pairing}

The basis of the description is SHF augmented by BCS pairing.  The method
is widely used and well documented in the literature, for reviews see
e.g. \cite{Ben03aR,Erl11aR}. We report here briefly the actual input
and usage. 

The mean-field state is a BCS state characterized by a set of
single-particle wavefunctions $\{\varphi_\alpha,\alpha=1...\Omega\}$
and corresponding BCS occupation amplitudes $v_\alpha$. The SHF energy
functional depends on density $\rho$, kinetic energy density $\tau$,
spin-orbit density $\mathbf{J}$, current $\vec{j}$, spin density
$\vec{\sigma}$, and spin kinetic density $\vec{\tau}$. We use the
standard form of the Skyrme functional, in some cases augmented
by an isovector spin-orbit term \cite{Rei95a}. We employ consistently
the full form including the time odd currents ($\vec{j}$,
$\vec{\sigma}$, $\vec{\tau}$) which play a crucial role in computing
the collective masses along the fission path (see section
\ref{sec:recipe}). For the pairing functional we use the density
dependent zero-range pairing force \cite{Kri90a,Dob01b} in the
stabilized form \cite{Erl08a} which reads in detail
\begin{subequations}
\label{eq:pairfun}
\begin{eqnarray}
  E_{\rm pair}^{\mathrm{(stab)}}
  &=&
  E_{\rm pair}^{\mbox{}} 
  \bigg(1-\frac {E_{\rm cutp}^{2}}{E_{\rm pair}^{2}}\bigg)
  =
  E_{\rm pair}^{\mbox{}} 
  -
  \frac {E_{\rm cutp}^{2}}{E_{\rm pair}},
\label{eq:stabfunc}
\\
  E_{\rm pair}
  &=&
  \frac{1}{4} \sum_{q\in\{p,n\}}V_\mathrm{pair,q}
  \int d^3r \xi^2_q
  \left[1 -\frac{\rho}{\rho_{0,\mathrm{pair}}}\right],
\label{eq:ep}
\\
  \xi_q
  &=&
 \sum_{\alpha\in{q}}f_\alpha{u}_{\alpha}v_{\alpha}|\varphi_{\alpha}|^2
 ,\;
 q\in\{\mathrm{prot,neut}\},
\end{eqnarray}     
\end{subequations}
where $u_\alpha=\sqrt{1-v_\alpha^2}$ is
the pair density and
$f_\alpha$ is a cut-off weight as defined in \cite{Ben00c}.
When proceeding along the deformation path, the pairing energy, 
$E_{\rm pair}$, is plagued by a possible phase transition 
to the breakdown of pairing which, in turn, leads to
singularities in the collective mass. A widely used method to
stabilize pairing against breakdown is to employ the Lipkin-Nogami
recipe, see e.g. \cite{Rei96a}. This, however, is not always effective
enough in the breakdown regime. The stabilized functional
(\ref{eq:stabfunc}) provides a much more robust scheme. We use it with
$E_{\text{cutp}}=0.6$ MeV 
which leaves ground state properties nearly unchanged and is at the
same time very efficient in suppressing singularities in the
collective mass.

The form of the SHF functional is more or less prescribed by a
low-momentum expansion of a fictitious underlying effective two-body
interaction \cite{Neg72a}. But the model parameters cannot yet be
derived from ab-initio methods with sufficient precision. 
It is customary then to 
adjust the parameters of the SHF functional
 to experimental data, mostly from ground state
properties \cite{Ben03aR,Erl11aR}. Steadily growing availability of
data from exotic nuclei and different preferences in the choice of the
fit data has led to a variety of SHF parameterizations. One needs to
check the results for sufficiently different parameterizations in
order to explore the predictive value of SHF calculations. We will
consider the following parameterizations:
SkM$^*$ \cite{Bar82a} as a traditional benchmark because it is one of
the first parameterizations delivering a quality description of
nuclear ground states;
SkI3 \cite{Rei95a} 
which for the first time exploits the freedom of an 
isovector spin-orbit coupling thus simulating in this respect the
situation in relativistic mean-field models;
SLy6 \cite{Cha98a}
which had been developed with a bias to neutron rich nuclei and neutron
matter aiming at astrophysical applications;
SV-min \cite{Klu09a} as a recent development using a large pool of
semi-magic nuclei which were checked to have negligible correlation
effects; 
and SV-bas which was adjusted to the same data as SV-min with an
additional constraint on nuclear matter properties (symmetry energy,
isoscalar and isovector effective masses) to tune giant resonances
together with ground state properties \cite{Klu09a}.
We will also show a result for HFB-14  as one 
representative in a large series of parameterizations derived in 
large scale fits biased on a comprehensive description of binding 
systematics, in this case referring to published data \cite{HFB14}.

All these parameterizations employed different pairing recipes in
their original definition.  In order to make calculations better
comparable, we use the same pairing functional (\ref{eq:pairfun}) for
all parameterizations and tune the pairing parameters
($V_\mathrm{pair,p}$, $V_\mathrm{pair,n}$, $\rho_{0,\mathrm{pair}}$)
to the data from even-odd staggering as summarized in \cite{Klu09a}.
Such a separate adjustment for each force is crucial because the
actual pairing gaps energy from an interplay of pairing strengths
$V_\mathrm{pair,q}$ and level density which depends sensitively on the
effective mass and thus varies dramatically with the parameterization
\cite{Ben03aR,Erl11aR}.

\subsection{Microscopic computation of fission lifetimes}
\label{sec:recipe}

Fission represents a substantial rearrangement of a nucleus from one
into two fragments. SCMF models are well suited to track this process
in a least prejudiced manner. They require only one constraint to
force a stretching of the system along the various stages while all
other details of the rearrangements and shapes result automatically of
the calculation. Actually, one uses an isoscalar quadrupole constraint
because the first stages of fission develop out of large-amplitude
quadrupole modes of the mother nucleus. Even this last piece of
guesswork could be eliminated by using the recipes of adiabatic
time-dependent Hartree-Fock (ATDHF) \cite{Rei87aR,Dan00aR}. This,
however, has not yet been accomplished for the very heavy systems
considered here. We stay at the level of constrained SHF and employ
ATDHF only for a self-consistent evaluation of the collective mass and
quantum corrections \cite{Rei87aR}.

Fission barriers have been discussed already in the early stages of
SCMF models and have even been used as benchmark for calibration
\cite{Bar82a}. The calculation of fission lifetimes are much more
involved as their computation requires not only the potential energy
surface along the fission path, but also the corresponding collective
masses and a safe estimate of the collective ground state correlations
for the initial state.  Thus, the vast majority of calculations of
fission lifetimes employ the microscopic-macroscopic method which
combines shell corrections with a macroscopic liquid-drop model
background, see e.g.  \cite{Mol87a,Smo95a}.  Self-consistent
calculations of fission lifetimes are still rare, see
e.g. \cite{Ber01a,War06}, and mostly use still approximate masses and
quantum corrections.  On the other hand, just because the computation
of fission lifetimes is so demanding they serve as extremely critical
observables probing all aspects of the effective nuclear interaction,
its global bulk properties as well as details of shell structure.

\begin{figure}
\epsfig{figure=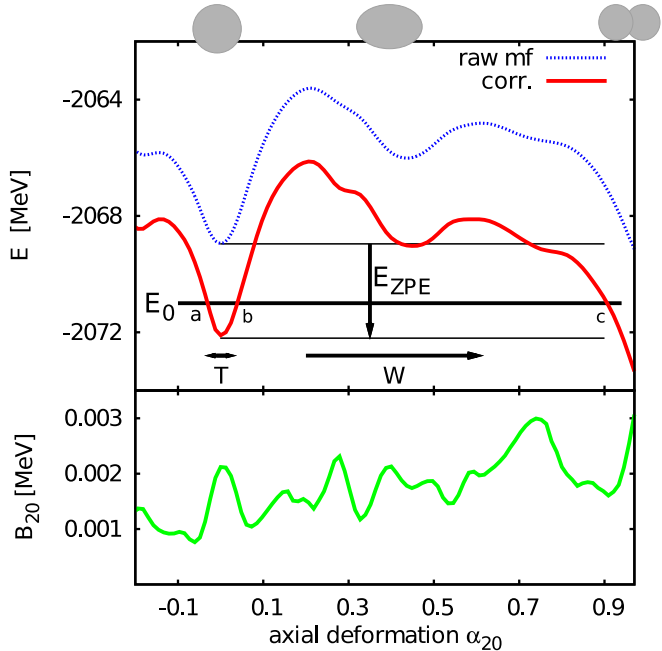,width=1.0\linewidth}
\caption{\label{fig:PES-fiss} {\it(Color online)}
  Upper panel: Raw potential energy
  surface (PES) $\mathcal{V}(\alpha_{20})$ and ZPE corrected PES
  ${V}(\alpha_{20})$ for $^{290}$Sg.  The energy $E_{0}$ of the
  collective ground state is indicated by a heavy horizontal line. The
  points $a$, $b$, and $c$ indicate the three crossing points of the
  $E_0$ line with the collective potential
  ${V}(\alpha_{20})$. Tunneling and repetition rates W and T are also
  indicated.  Lower: Inverse collective mass $B_{\alpha_{20}}$
  calculated by a self-consistent cranking scheme (ATDHF
  cranking). Schematic nuclear shapes are sketched.}
\end{figure}

To calculate fission lifetimes within SHF we use the scheme
as presented in \cite{Sch09a}.  We consider a fission path
which evolves along axially symmetric shapes and characterize the
deformation by the dimensionless axial quadrupole momentum
\begin{subequations}
\begin{eqnarray}
   \alpha_{20}
   &=&
   \frac{4\pi}{5}
   \frac{\langle\Phi|\hat{Q}_{20}|\Phi\rangle}{\langle\Phi|r^2|\Phi\rangle}
   \quad,
\\
   \hat{Q}_{20}
   &=&
   \sqrt{\frac{5}{16\pi}}\left(2z^2-x^2-y^2\right)
   \quad.
\end{eqnarray}
\end{subequations}
The steps to compute the fission lifetimes can then be summarized as:
\begin{enumerate}  
\item The fission path is a set of mean-field states
  $\{|\Phi_{\alpha_{20}}\rangle\}$ representing the stages on  the way from the
  ground state to fission. It is
 generated by quadrupole-constrained SHF 
(complementing the mean-field Hamiltonian by a constraining potential, i.e., $\hat{h}
\rightarrow \hat{h} - \boldsymbol\lambda \cdot {\hat{\bf Q}}_{20}$).
\item
The 
energy expectation value corresponding
to $|\Phi_{\alpha_{20}}\rangle$ yields a ``raw'' collective energy
surface, 
$\mathcal{V}(\alpha_{20})$.
\item The collective mass, $B(\alpha_{20})$, and moments of inertia
  are computed by self-consistent cranking (often called ATDHF
  cranking) along the states $|\Phi_{\alpha_{20}}\rangle$ of the path
  \cite{Klu08a}.
\item Approximate projection onto angular momentum zero is performed using
the moments of inertia and angular-momentum width.
\item Quantum corrections for the spurious vibrational zero-point
  energy (ZPE) are applied (using quadrupole mass and width). The
  result is the ZPE corrected potential energy surface (PES),
  ${V}(\alpha_{20})$.
\item\label{it:gs} 
The collective ground state energy $E_{0}$ is computed fully quantum
mechanically for the thus given collective Hamiltonian \cite{Klu08a}.
\item The tunneling rate $W$ at the given ground state energy and the
  repetition rates $T$ are computed by the standard semi-classical
  formula (known as WKB approximation) using the
  quantum-corrected potential energy and collective mass (moments of
  inertia); the fission lifetime is finally composed from these two
  quantities as $T/W$:
\begin{eqnarray} 
\label{equ:period-koll}
  W
  &=&
\exp{\left(-2\int_b^c\sqrt{\frac{V(\alpha_{20})-E_0}{B(\alpha_{20})}} d\alpha_{20}\right)}
\\
  T
  &=&
\hbar\int_a^bd\alpha_{20}
\left({\sqrt{{B(\alpha_{20})}({E_0}-{V(\alpha_{20})})}}\right)^{-1}
  \quad ,
\label{eq:tunnel-koll}
\end{eqnarray}
where the initial state would be classically bound in the interval $(a,b)$,
while the barrier extends over the interval $(b,c)$ (see figure 
\ref{fig:PES-fiss}).
\end{enumerate}
Point \ref{it:gs} in this list requires some explanation.  The axially
symmetric fission path is described by three collective degrees of freedom
(deformation $\alpha_{20}$ and two rotation angles) while the full collective
quadrupole dynamics calls for the five-dimensional Bohr Hamiltonian.  In order
to be consistent with the whole fission path, a three-dimensional collective
dynamics for quadrupole motion was derived which is restricted to axially
symmetric shapes.  The method employs the norm and overlap kernel of the
topological Gaussian overlap approximation~\cite{Rei78b,Hag02a}
and relies on the direct solution of the
collective Schr\"odinger equation (see \cite{Klu08a}), rather than
establishing a (reduced) Bohr Hamiltonian.

Figure \ref{fig:PES-fiss} illustrates the collective parameter
functions along the axially symmetric path for the super-heavy element
$^{290}$Sg which are necessary for calculation of the fission
half-lives.  As can be seen in figure \ref{fig:PES-fiss} the
collective mass fluctuates strongly so it can hardly be approximated
by some constant or weakly changing collective mass.

As mentioned above, all calculations are performed in axial symmetry but
allowing for reflection asymmetric shapes. This breaking of reflection
symmetry becomes crucial in the outer region beyond the fission barrier. The
ground states and the (first) barrier are usually associated with reflection
symmetric shapes with few exceptions as discussed 
in the appendix. Breaking of axial symmetry towards triaxial shapes can
occur in the barrier region.  One knows from actinides that triaxial shapes
can lower the barriers by about 0.5--2 MeV \cite{Cwi96a,Ben98a}. Such lowering
is missing in axially symmetric calculations.  The present results are thus to
be understood as providing an upper limit for barriers and lifetimes.

\section{Results and discussion}
\label{sec:results}

\subsection{Benchmark}

\begin{figure}
{\epsfig{figure=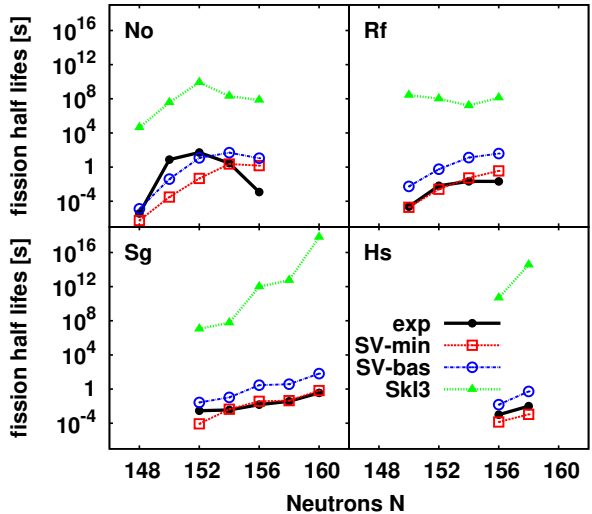,width=1.22\linewidth}}
\caption{\label{fig:hlexp} {\it(Color online)}
  Fission lifetimes of the isotopes No, Rf,
  Sg and Hs computed with three different Skyrme parameterizations, as
  indicated, and compared with data from
  \cite{Aud03,Khu08,Pet06,Oga01b,Gat08,Greg06,Dvo08,HofS01}} 
\end{figure}

Lifetimes can be {derived} by the calculation scheme developed
above.  Figure \ref{fig:hlexp} shows fission lifetimes for four chains
at the lower end of SHE. Three SHF parameterizations are
compared. SkI3 which has a rather low effective mass ($m^*/m=0.58$)
does not perform so well. This holds similarly for all forces with low
effective masses.  The results from the recent parameterizations
SV-min and SV-bas which have effective mass $m^*/m=0.95$ and $0.9$
provide a more satisfying agreement taking into account that an order
of magnitude description is already a success for the extremely subtle
observable of fission lifetime.  For the No isotopes, the agreement is
acceptable in the average but there appears a strong deviation in the
isotopic trend. This mismatch stems probably from the axial
approximation and could well be explained by a strong isotopic change
of the triaxial lowering of the barrier.

\begin{figure}
\centerline{\epsfig{figure=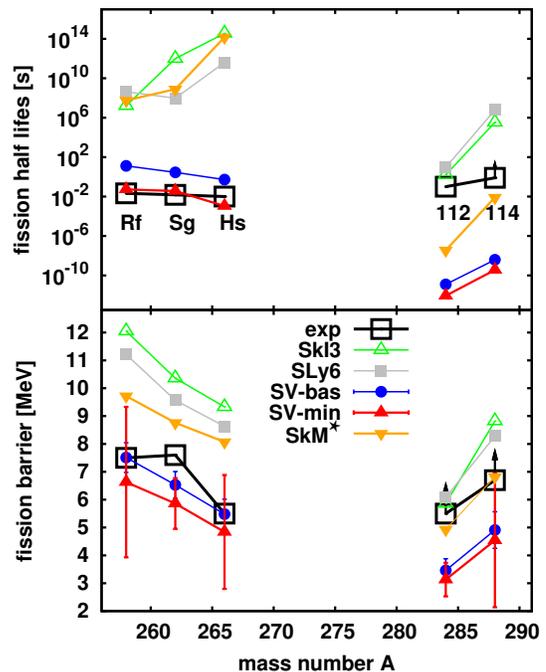,width=0.85\linewidth}}
\caption{\label{fig:hlbarexp} {\it(Color online)}
  Fission barriers (lower) and life-times
  (upper) for two groups of experimental known super-heavy
  elements. Compared are results from a variety of Skyrme
  parameterizations with experimental data
  \cite{Aud03,Hof01,Tan06a,Oga06,Itk02,Pet04}.
  The error bars on the barriers for SV-min are the uncertainties in
  the extrapolation as implied in the least-squares fits of the
  parameterization \cite{Klu09a}.
}
\end{figure}

Figure \ref{fig:hlbarexp} shows experimental and calculated results on
fission barriers (lower panel) and lifetimes (upper panel) for a few
selected SHE, but now extending to heavier elements and comparing more
SHF parameterizations.  The SHE represent two groups, one
at the lower side (already included in figure \ref{fig:hlexp}) and
another one with much heavier nuclei at the limits of present days
data.  The span of predictions from the various Skyrme forces is huge
in all cases in spite of the fact that all these parameterizations
provide a high-level description of nuclear ground state properties
along the valley of stability.  The variation of predictions may be a
welcome feature as it provides additional selection criteria for a SHF
parameterization. There remains, however, a problem when looking at
the trend from the lighter side (Rf, Sg, Hs) to the heavier elements
(Z=112, 114). All parameterizations produce a wrong trend of the
predictions from the lower to the upper region.  Barriers and
lifetimes are well reproduced in the lower group by SV-min and
SV-bas. But these parameterizations underestimate the barrier heights
and lifetimes for the upper group \cite{Erl10a}.  The problem persists
even with a more flexible density dependence of the Skyrme functional
\cite{Erl10b}. It is also unlikely that triaxiality, ignored here,
could help. It would worsen the situation for SV-min and SV-bas and
the possible lowering about 0.5--2 MeV is insufficient to bridge the
gap for the other parameterization.  One has to keep in mind, however,
that an experimental determination of lifetimes and barriers for the
heaviest elements is a very demanding task and the data may not yet
have reached their final stage such that the mismatch should presently
not be over-interpreted.  In any case, we can expect from modern
parameterizations as, e.g., SV-min a pertinent picture of the
systematics of fission lifetimes for SHE.

\subsection{Fission topologies}

\begin{figure*}
\centerline{\epsfig{figure=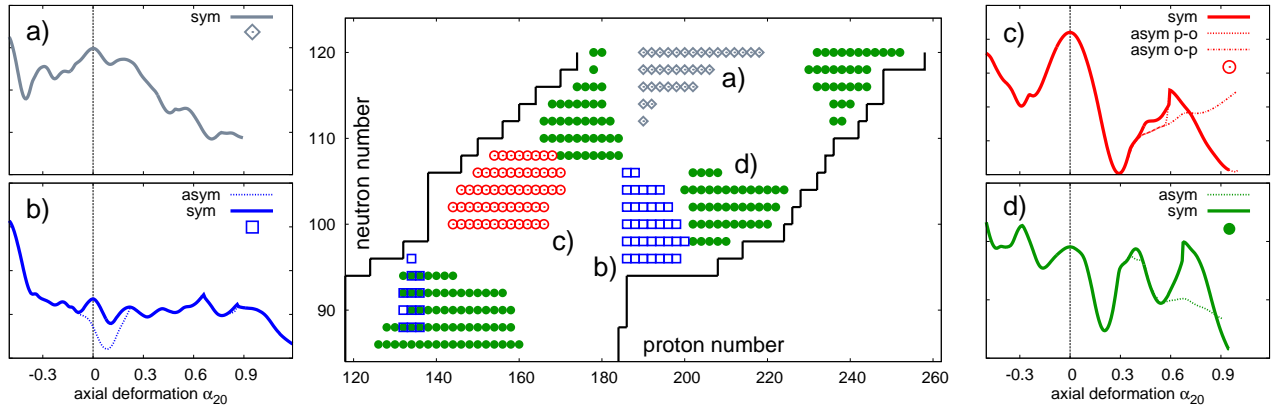,width=1.0\linewidth}}
\caption{\label{fig:schematic} {\it(Color online)}
  Schematic survey of fission modes for
  the various SHE. The two panels on the left and right show examples
  for the different regions as indicated: a) triaxial fission b)
  asymmetrical ground-state energy c) bi- or multi-modal fission d)
  asymmetric calculation removes completely the outer barrier.  }
\end{figure*}

When going through the variety of SHE, one encounters much different shapes of
the fission landscape.  Four different situations can be distinguished as
shown schematically in figure \ref{fig:schematic}:
\begin{description}

\item[Panel a] shows a case where the potential energy surface has a strong
  oblate minimum and optionally a secondary prolate minimum.  It is thus
  assumed that the fission path is going through the triaxial
  plane \cite{Ben01d}.
  In fact, we have seen
  from fully triaxial calculations that in most cases already the ground state
  acquires some triaxiality.  Therefore, it is not possible to make reliable
  predictions using an axial code in such cases.  No fission barriers and
  lifetimes will be shown for this region in the following sections.

\item[Panel b] shows a PES for the case of a reflection asymmetric ground
  state which occurs in two region in $88 \le Z \le 96$ and $132 \le N \le
  136$ and again in $96 \le Z \le 108$ and $186 \le N \le 198$.  The breaking
  of reflection symmetry enhances the binding which, in turn, increases the
  fission barrier and thus leads to enhanced fission lifetime. This aspect
  will be discussed in more detail 
in the appendix.

\item[Panel c] shows a case where different fission paths emerge depending on
  whether one restricts the calculation on reflection symmetry or not or
  whether the calculation is going from oblate to prolate deformation or vice
  versa.  This suggests a structure of two valleys where symmetric fission
  competes against asymmetric fission. This is also called multi- or bimodal
  fission and was already discussed in detail for self-consistent mean field
  models \cite{Ben98a,Sta09a,War02a} as well as for the
  microscopic-macroscopic finite-range liquid-drop model \cite{Moe09} (and
  citations therein). A full multi-modal treatment is presently beyond our
  possibilities. But the figure indicates that the barriers in the different
  channel are not so dramatically different. Thus it still provides a
  pertinent picture if we consider one particular path, the one along
  asymmetric shapes 
starting from outside.

\item[Panel d] shows the standard case which has one unique barrier in
  asymmetric calculations. In order to demonstrate the effect of asymmetry we
  compare with the PES from reflection symmetric calculations. The latter show
  the double-humped fission barrier as it is known from actinides
  \cite{Bjo80a}. The allowance of asymmetric shapes removes the second barrier
  which holds for practically all SHE \cite{Bue04a}.

\end{description}
The overview demonstrates the large
variety of topologies for the fission PES. This inhibits an automatic barrier
search. Thus for the greater part of the investigated PES the minima and
maxima relevant for the fission process were determined manually.

\subsection{Systematics of fission barriers}

\begin{figure}
\centerline{\epsfig{figure=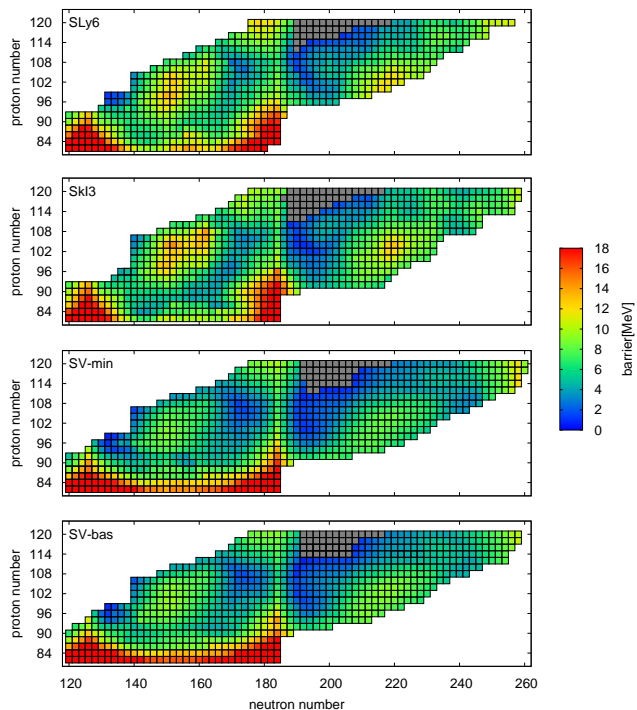,width=\linewidth}}
\caption{\label{fig:sysbarr1}{\it(Color online)}
Systematics of fission barriers of even-even nuclei for four different Skyrme 
parameterizations SLy6 \cite{Cha98a}, SkI3 \cite{Rei95a}, SV-min \cite{Klu09a} 
and SV-bas \cite{Klu09a}. Grey indicates a possible triaxial fission, 
as already seen in figure \ref{fig:schematic}.
}
\end{figure}

A simple first indicator of fission instability is the height of the fission
barrier.  Figure \ref{fig:sysbarr1} shows the systematics of fission barriers
of SHE in the range $82\le Z\le 120$ and $120 \le N \le 260$.  All elements
are found to be stable against immediate nucleon emission at the ground state
and along the whole fission path.  In case of a double-humped structure of the
barrier (commonly appearing for $Z < 100$) the higher barrier is plotted.
Results are shown for the four different Skyrme parameterizations SLy6, SkI3,
SV-min and SV-bas.

At first glance it is apparent that the Skyrme forces SkI3 and SLy6 yield
notoriously much higher barriers as SV-min and SV-bas. This can be traced
back to a difference in the effective mass $m^*/m$. SLy6 and SkI3 have a very
low mass $m^*/m=0.69$ and $m^*/m=0.58$, respectively while SV-bas and
SV-min have effective masses 0.9 and 0.95.  A low effective mass leads to a
too low density of single-particle states, and thus to larger 
shell correction energies which, in turn, yield larger barriers.

All forces provide the same trends over the landscape of SHE.  There is a
strong variation in fission barriers corresponding to the strong variations of
shell structure in the landscape of SHE. Several regions of high barriers
occur.
For low $Z$, i.e. Rn ($Z$=86), Ra ($Z$=88), and Th ($Z$=90), one
sees two islands of high barriers, one at $N\approx126$ and another
one at $N\approx 184$. Neutron numbers in between cover
a region of lower barriers. They are particularly low for SkI3 and
to some extend SLy6 while the fluctuations between high and low
barriers are less dramatic for SV-bas and SV-min, corresponding to
their generally smaller shell corrections.
  Stepping up to higher proton numbers $Z$ there follow two more
regions of high barriers, one of deformed SHE around $Z/N
=104/152$ and one of spherical SHE around $Z/N=120/184$ (spherical shell
closure).  The magic neutron number N=184 is clearly visible while an expected
magic proton number near $Z=120$ is indicated by a broad island of enhanced
stability barriers around $Z/N=120/184$ \cite{Ben01a}.

Considering the regime of nuclei relevant for r-process
nucleosynthesis it is interesting to notice the appearance of a region
of low fission barriers for $Z\sim 84$ in moving from $N=126$ to
$N=184$ for SLy6 and particularly for SkI3. These low fission barriers
may allow for neutron-induced fission to occur as the nucleosynthesis
flow moves from the $N=126$ region to the $N=184$. The situation is
different for SV-min and SV-mass, that predict much larger fission
barriers in the r-process relevant region. For these parameterizations
fission will only be relevant once the nucleosynthesis flow overcomes
the $N=184$ magic number. All four forces agree in predicting a rapid
decrease of barrier heights going from the shell closure $N=184$ up to
neutron-rich nuclei. This suggest a substantial decrease in the
production of nuclei beyond $N=184$ during the r process. These
aspects will be explored in a forthcoming publication.

\begin{figure}
\centerline{\epsfig{figure=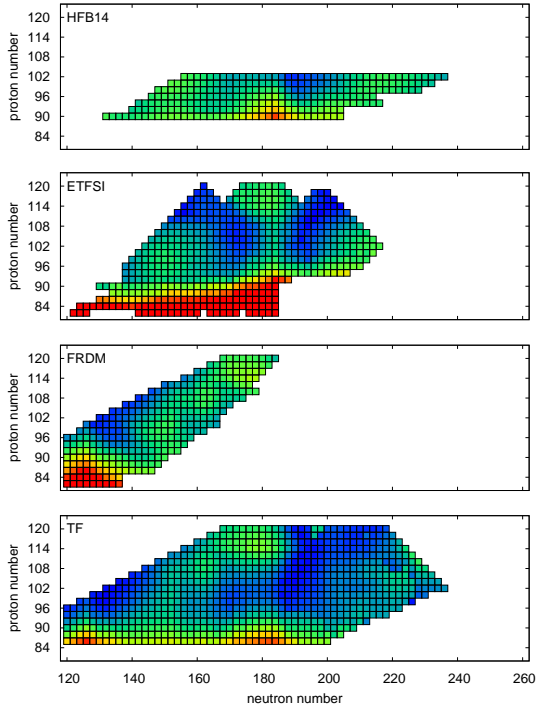,width=\linewidth}}
\caption{\label{fig:sysbarr2}{\it(Color online)}
Results for fission barriers of four theoretical mass models HFB-14 \cite{HFB14}, ETFSI \cite{ETFSI}, 
FRDM \cite{Moe09}, TF \cite{Mye99a} in the same region of nuclei. The shown data are reduced to 
even-even nuclei to enable a better comparison.
}
\end{figure}

To complete the picture, figure \ref{fig:sysbarr2} shows results from
four other models for which data are publicly available.  The
calculation were performed by using the Extended Thomas-Fermi plus
Strutinsky integral (ETFSI) \cite{ETFSI}, the Thomas-Fermi (TF)
\cite{Mye99a} method , the macroscopic-microscopic finite range
liquid-drop model (FRDM) \cite{Moe09}, or the SHF approach with the
parameterization HFB14 \cite{HFB14}.  All four theoretical mass models
shows similar trends {as observed in our studies}.  They confirm the
region of high barriers below Uranium, around neutron number $N=184$
and the island around $Z/N = 104/152$, the latter, however, less
strongly developed in case of TF and FRDM.  The three non
self-consistent models (TF, ETFSI, FRDM) shift the third island of
stability ($Z/N=120/184$) down towards proton number $N=114$.  The
rapid fall-off beyond $N=184$ is confirmed.  The case HFB14 belongs to
the SHF family.  In spite of the narrow range of results one can
conclude that its systematics is very similar to the other SHF
cases. The actual barrier height are closer to SV-min and SV-bas
(sometimes even below). This is not surprising as all models in
figure~\ref{fig:sysbarr2} have a large effective mass, equals or
around $m^*/m=1$.

\subsection{Fission lifetimes}

\begin{figure}
{\epsfig{figure=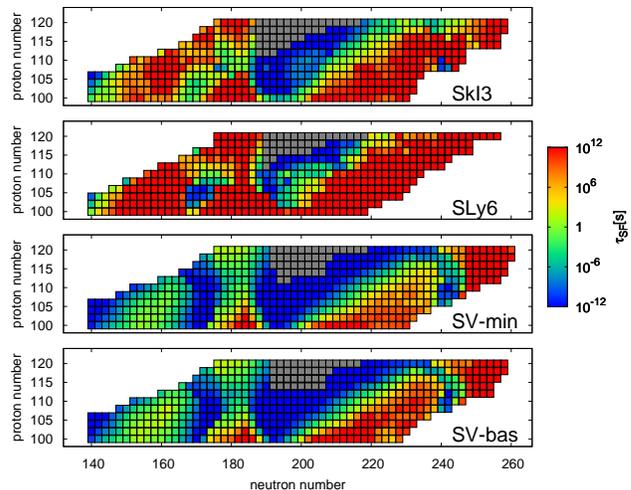,width=\linewidth}}
\caption{\label{fig:syshl}{\it(Color online)}
Systematics of fission lifetimes calculated in the range of proton numbers $100\le Z\le 120$ and 
neutron numbers $140 \le N \le 260$ for even-even nuclei using the Skyrme parameterizations 
SkI3, SLy6, SV-min and SV-bas.}
\end{figure}

Figure \ref{fig:syshl} shows fission lifetimes calculated with the recipe as
outlined in section \ref{sec:recipe} in the range $100\le Z\le 120$ and $140
\le N \le 260$ (from proton to neutron drip line) for the same Skyrme
parameterizations as in the barrier systematics.  Calculation of lifetimes
were not performed for nuclei with proton numbers $Z\le98$, because the
doubled humped barrier makes the evaluation of lifetimes cumbersome.  The
difference in barrier heights from 0 to 12 MeV (figure \ref{fig:sysbarr1})
translates to a difference in lifetimes from almost immediate decay to
$10^{12}$ s and longer, demonstrating again the enormous sensitivity of
fission lifetimes to all details of the model and computation.

At first glance, the basic pattern resemble much the systematics of
barriers.  Long lived SHE are obviously found in the islands of high
barriers.  The island around $Z/N=104/152$ is even broadened to higher
$Z$ and $N$ towards the assumed neutron shell closure at N=162,
especially for SV-min and SV-bas.  This demonstrates that not only
barrier but also barrier width and collective mass can have a decisive
influence.  All parameterizations show a broad and deep valley of
fission instability starting abrupt with neutron number $N=186/188$.
If the r-process nucleosynthesis flow is able to overcome the $N=184$
magic number, it will proceed by the region of large spontaneous
fission lifetimes in figure~ \ref{fig:sysbarr1}. However, once the
neutrons are exhausted and matter beta-decays, the region of short
spontaneous fission lifetimes will be reached and no long live SHE
will be produced. The situation may be different for SkI3 and SLy6,
depending on the extend of the region of short lifetimes above
$Z>120$.

There is a large difference between the SHF parameterizations in
overall lifetime for elements with $N<190$.  The nuclei are much more
stable for SkI3 and SLy6 than for SV-bas and SV-min.  This is, of
course, related to the overall difference in barrier heights (see
figure \ref{fig:sysbarr1}) which can be traced back to different
effective masses $m^*/m$.  This produces here even a qualitative
difference: The parameterizations SV-min and SV-bas show a valley of
fission instability between the islands around $Z=120$, $N=180$ and
$Z/N =104/152$ while SkI3 and SLy6 do not. The immediate consequence
is that SkI3 and SLy6 predict uninterrupted chains of $\alpha$ decay
from the heaviest SHE down to actinides while SV-min and SV-bas have
these $\alpha$ chains terminated by spontaneous fission. The latter is
what is empirically found \cite{Oga06}. The
competition with $\alpha$ decay is discussed in section
\ref{sec:alpha}.

\subsection{$\alpha$-decay}
\label{sec:alpha}

The $\alpha$-decay half-lives are evaluated 
using the Viola-Seaborg relationship \cite{Vio66a,Sob89a}:
\begin{eqnarray}
\log(\tau_\alpha/\text{s})&=&(a Z+b)(Q_\alpha/\text{MeV})^{-1/2}\\
&&\!+(c Z+d)+h_{\log} \\[0.5cm]
a&=&1.66175,\quad\; b=-0.5166,\\ c&=&-0.20228, \;d=-33.9069,
\end{eqnarray}
\begin{eqnarray}
  Q_\alpha
  &=&
   E(N\!-\!2,Z\!-\!2)+E(2,2)-E(N,Z) 
  \;,
\\
   E(2,2)
   &=&
   E_{\text{exp}}(^4\text{He})   =   28.3\ \text{MeV}
   \;,
\end{eqnarray}
and $h_{\log}=0$ for the even-even nuclei considered here.
This requires as only input the $Q_\alpha$ values which can be
determined easily as difference of ground state binding energies.  The
latter are computed allowing for axial deformations as well as
reflection-symmetry breaking and including approximate angular
momentum projection for deformed nuclei. It is to be noted that the
difference of binding energies, as the $Q_\alpha$ value, are predicted
rather reliably with the SHF models although the binding energies as
such are notoriously underestimated for SHE \cite{Klu09a,Erl10a}.

\begin{figure}
{\epsfig{figure=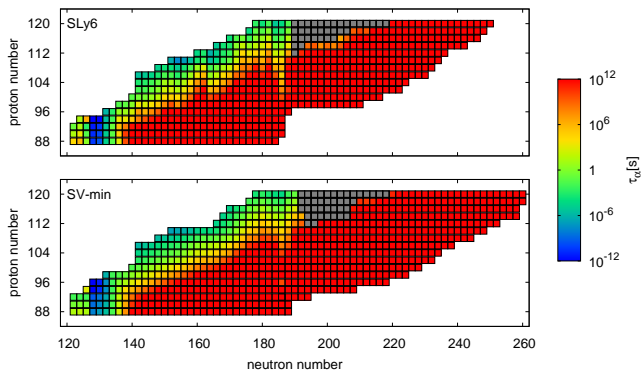,width=\linewidth}}
\caption{\label{fig:syshlalp2}{\it(Color online)}
$\alpha$-decay half-lives computed by using the $Q_\alpha$ values 
and the semi empirical Viola Seaborg formula. 
}
\end{figure}

Figure \ref{fig:syshlalp2} shows the systematics of $\alpha$-decay
half-lives calculated with the SHF parameterizations SLy6 and SV-min.
In contrast to fission lifetimes, $\alpha$-decay half-lives vary in
general smoothly and steadily with a tendency to increase when going
in direction of neutron rich SHE. An exception are the spherical
neutron shell closures at N=126 and N=184 which are clearly marked by
a sudden decrease of $\alpha$ half-lives. But there is no detailed
structure like the islands of stable nuclei in spontaneous fission
systematics.  Most of the nuclei in the shown region are very stable
against $\alpha$ decay. It is only the band of neutron-deficient SHE
at the left side of the region where $\alpha$ decay plays a role as
competitor to fission and $\beta$ decay. Comparing with the fission
lifetimes in figure \ref{fig:syshl} that $\alpha$ decay prevails in
any case for the islands of fission stability around $Z=120$, $N=180$
and $Z/N =104/152$. The parameterizations SV-min and SV-bas produce
the pronounced valley of fission instability between these islands for
which then fission takes the lead over $\alpha$ decay. This does not
happen for SLy6 and SkI3 with their generally longer fission life
times.

\begin{figure}
\centerline{\epsfig{figure=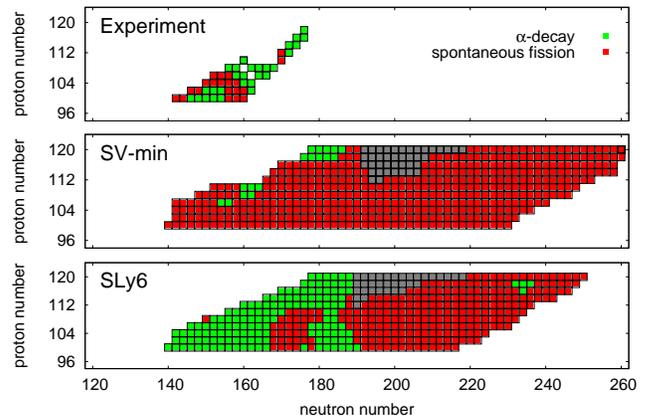,width=1.0\linewidth}}
\caption{\label{fig:minlt}{\it(Color online)}
Minimal lifetime plots for the competing decay channels $\alpha$-decay (green), 
and spontaneous fission (red). Experimental known decay channels
\cite{Aud03} are compared to results using the parameterization 
SV-min and SLy6.}
\end{figure}

A direct comparison of $\alpha$ decay and fission is provided in
figure \ref{fig:minlt} showing the systematics {of} the
dominant decay channels.  Results are shown for the two
parameterizations SV-min and SLy6 and compared with {data}. The
experimental situation is just in between the two theoretical
predictions. SLy6 produces too much fission stability thus giving
$\alpha$ decay too much dominance while SV-min slightly underestimates
the impact of $\alpha$ decay.  Consider the $\alpha$ decay chain from
$Z=118$, $N=176$. It is terminated by fission already for
${Z=116}$ for
SV-min while the experimental chain terminates later at $Z=112$.  The
results for SkI3 are very similar to those of SLy6 and the results for
SV-bas to those of SV-min.

It is to be noticed that the trends which are seen here in the
systematics had already been spotted in figure \ref{fig:hlbarexp}.
The parameterizations SV-min and SV-bas perform generally better, but
are still plagued by producing a wrong trend when stepping to the
heavier end of SHE where they yield {too} low barriers and lifetimes.
A better compromise has yet to be worked out.

\section{Conclusions}

We have explored the systematics of fission barriers and lifetimes in
the realm of super-heavy elements (SHE) on the grounds of
self-consistent calculations using the Skyrme-Hartree-Fock (SHF)
approach.  The fission path has been generated with a quadrupole
constraint producing a series of axially symmetric deformations while
allowing for reflection-asymmetric shapes.  The corresponding
collective mass is computed by self-consistent cranking (often called
ATDHF cranking). The quantum corrections to the collective potential
(angular momentum projection, vibrational zero-point energy) are
properly taken into account.  The fission life-time is computed for
thus given potential and mass by the semi-classical WKB approximation,
while the ground state energy, which is at the same time the entrance
energy for fission, is computed quantum mechanically in the given
collective geometry of one axial deformation plus {two}
rotation angles.  Results have been produced for a couple of different
SHF parameterizations to explore the sensitivity to the
parameterization.  For comparison, we have also computed the
$\alpha$-decay life-times using the Viola systematics.

A first test was performed by comparing with known fission life-times
in isotopic chains in the lower region of super-heavy elements,
$Z=104$--108 and for the few available data points in even heavier
elements. The span of predictions is large whereby the effective mass
of the underlying parameterization plays a crucial role. Satisfying
agreement is found for modern parameterizations using effective mass
around $m^*/m=0.9$--1. There remains, however, one open problem
with the global trend:
The parameterizations which perform almost perfectly in the
region $Z=104$--108 underestimate fission barriers and lifetimes in the
heavier region $Z=112$ and 114.

The landscape of SHE separates into regions of different topology of
the fission path. The most widely found standard case is a unique,
axially symmetric fission path showing only one fission barrier; the
second barrier which is {known} from actinides is suppressed by
reflection asymmetric shapes which regularly develop for larger
deformations. Proton rich isotopes in $100\leq Z\leq 108$ show often a
tendency to bi- or multi-modal fission where different fission and
fusion paths compete. A small region around $Z=118$ and $N=200$ has
oblate (if not triaxial) ground states and can decay only through a
manifestly triaxial fission path. There are, furthermore, two small
regions where the ground state is reflection asymmetric. This was
shown to enhance barrier and lifetimes at a quantitative level, but
not changing the global trends.

The systematics of fission barriers and lifetimes shows the known
islands of stability around $Z/N = 104/152$ in the region of a
deformed shell closure and around $Z/N = 118/178$ in a region
{of} spherical isotopes. The actual values of barriers and
lifetimes depend very much on the SHF parameterizations. Those with
low effective mass (here SkI3 and SLy6) produce very high barriers and
lifetimes while those with high effective mass (SV-min and SV-bas)
yield moderate barriers and lifetimes. The latter group also produces
a valley of fission instability between the two islands, qualitatively
in accordance with the empirical findings. The way to very neutron
rich $r$-process nuclei with $N>184$ starts out with a large region of
fission instability. Some stability is gained at the extremely neutron
rich end. This makes unlikely the production of {long-lived}
SHE above $Z=100$ by the r process. However, a more realistic estimate
of the production of SHE by the r process will require nucleosynthesis
calculations based on the present barriers and lifetimes. This will be
the subject of a forthcoming publication.

We have also computed $\alpha$-decay lifetimes using the Viola-Seaborg
formula.  While the fission life-times show dramatic variation over
the chart of super-heavy elements (from instability to
$\tau_\mathrm{fiss}=10^{16}$~s), the $\alpha$-decay times vary gently
with small overall changes and without visible shell effects.  The
general crossover from $\alpha$-decay to fission along the decay
chains from the upper island of SHE is qualitatively reproduced by the
family of SHF parameterizations with high effective mass.  A
quantitatively reliable prediction of the switching point is detail on
which the models have yet to be refined.

\appendix*
\section{Symmetric vs. Asymmetric (incl. isomer) \label{sec:symvsasym}}

Reflection asymmetric shapes are a key issue in fission of SHE and
thus have been much debated under different aspects as, e.g.,
suppression of the second barrier, impact of ground state asymmetry on
the first barrier, or influence on bi- and multi-modal fission
\cite{Moe09,Bon06a,War02a,Sta09a}.  We will discuss here the effect
of reflection asymmetric ground states on  the fission barrier and
subsequently on lifetimes.

\begin{figure}
{\epsfig{figure=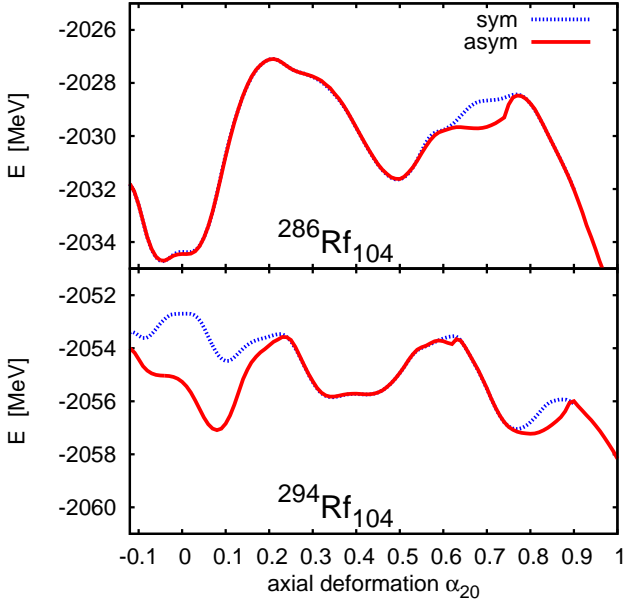,width=\linewidth}}
\caption{\label{fig:PES-sym-asym}{\it(Color online)}
PES of the isotopes $^{294}$Rf and $^{294}$Rf for symmetric and asymmetric shapes 
calculated with the Skyrme parameterization SLy6.}
\end{figure}

Figure \ref{fig:PES-sym-asym} shows the PES of the two isotopes
$^{286}$Rf and $^{294}$Rf. 
In the majority of the cases symmetric and asymmetric calculations
provide the same ground state energy. This is illustrated by
$^{286}$Rf. In contrast, for $^{294}$Rf the asymmetric calculation
yields an energetically more favorable ground state (around
$\alpha_{20}=0.08$), while in the symmetric PES it is difficult to
locate the minimum (possibly around $\alpha_{20}=0.32$).  It is to be
remarked that the tendency to symmetry breaking is confined to the
ground state region thus lowering the ground state energy. The
absolute height of the barrier remains almost unaffected. As a
consequence, an asymmetric ground state will lead to higher (relative)
fission barriers.

\begin{figure}
\centerline{
{\epsfig{figure=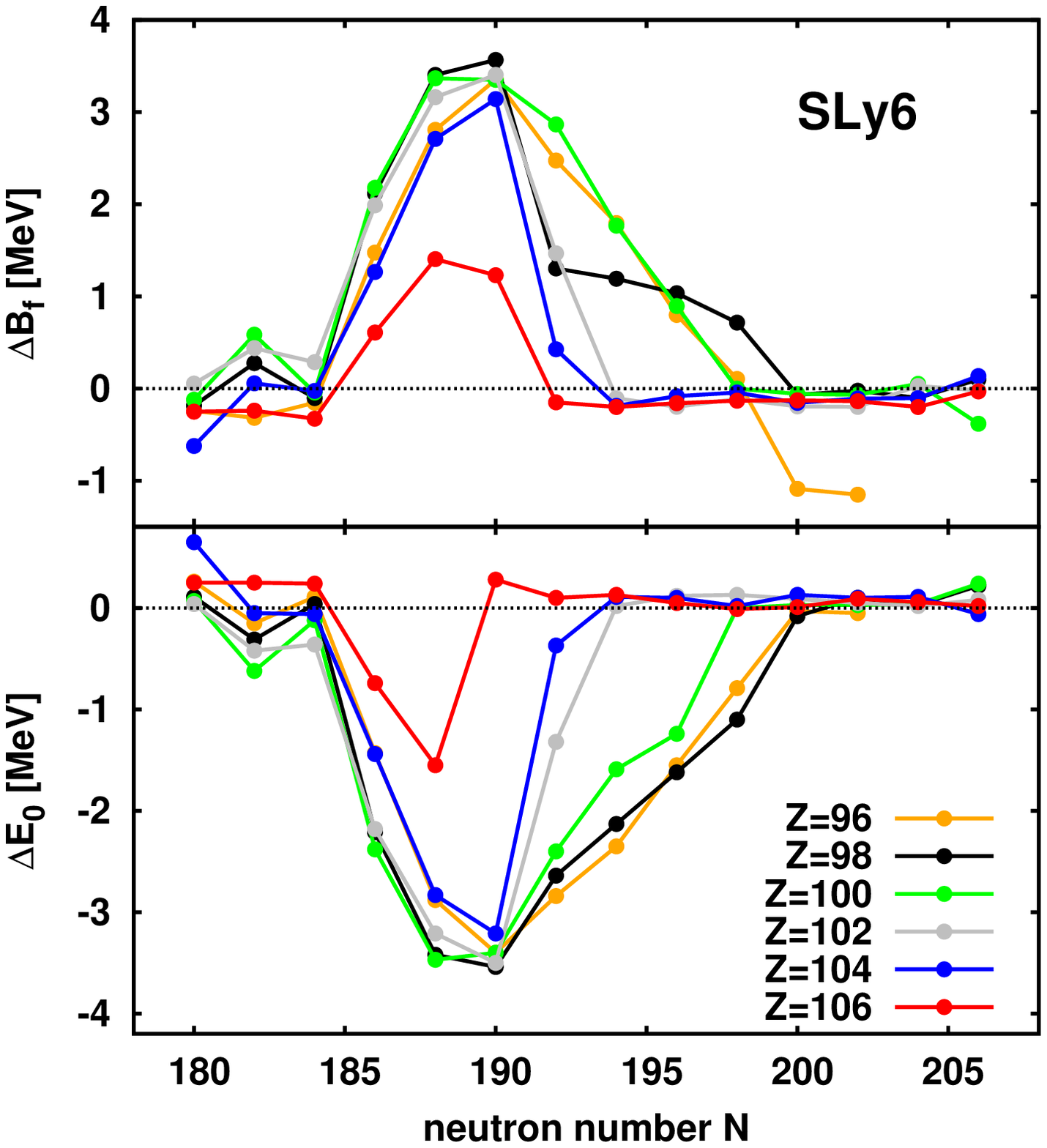,width=0.85\linewidth}}
}
\centerline{
{\epsfig{figure=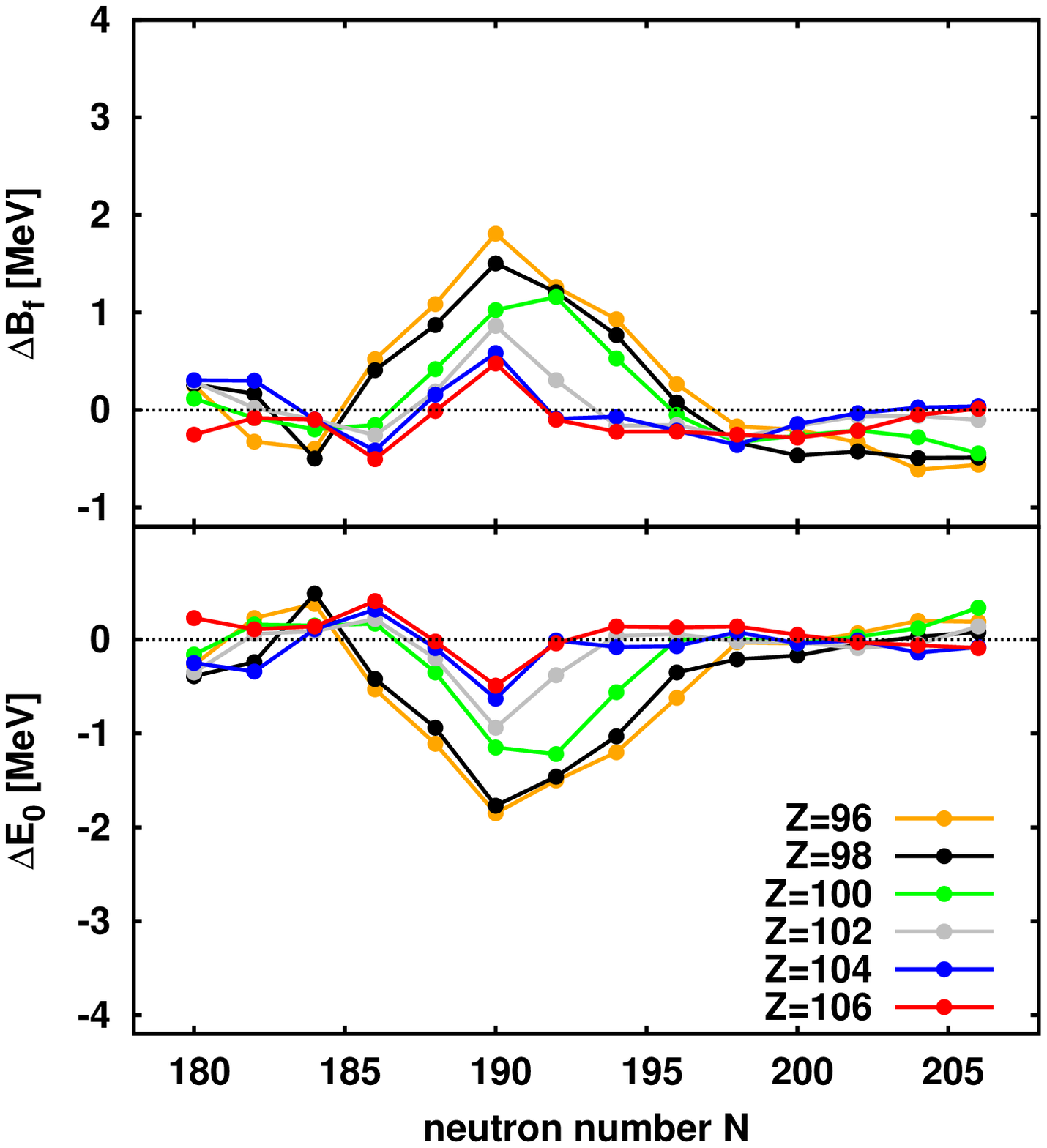,width=0.85\linewidth}}
}
\caption{\label{fig:Eground-sym-asym}{\it(Color online)}
Difference of ground state energies $\Delta E_0$ and fission barriers $\delta B_f$
between reflection symmetric and asymmetric calculations 
for the isotopes as indicated.
The upper panels show results computed with 
the Skyrme parameterization SLy6 and the lower panels with SV-min.
}
\end{figure}

Besides Rf, the elements Pu, U, Th, R and Rn are also known for the
importance of the octupole degree of freedom (\cite{Naz84a} and
\cite{Egi91a}).  Figure \ref{fig:Eground-sym-asym} summarizes for all
relevant elements the differences of binding energies and fission
barriers between reflection symmetric and asymmetric calculations for
SLy6 (upper panels) and SV-min (lower panels). All these isotopes
display basically the effect as discussed for Rf, namely that the
asymmetric shape affects predominantly the ground state thus leading
to an increase in the fission barrier which corresponds directly to
the lowering of the ground state. The size of the effect changes
quickly with proton and neutron number which is no surprise because
symmetry breaking is driven by (quickly changing) shell structure.
This also explains that the results from SLy6 and SV-min are
quantitatively so much different. SLy6 has a significantly lower
effective mass than SV-min thus lower level density and, in turn,
larger shell corrections.  The lowering of the ground-state energy was
also investigated using other Skyrme forces \cite{LoeDiss11}, where
the results show a strong dependence on the effective mass $m^*/m$ and
the pairing strength. A small effective mass or a small pairing
strength lead to a big effect on the ground state energy and vice
versa.

\begin{figure}
{\epsfig{figure=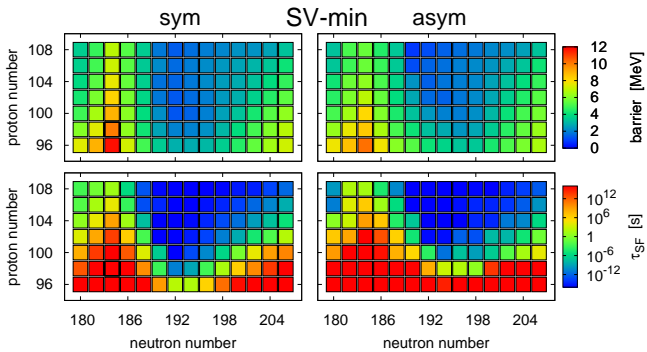,width=\linewidth}}
\caption{\label{fig:sf-barrier-sym-asym}{\it(Color online)}
Comparison of symmetric and asymmetric fission barriers and half-lives  
in region b) of the schematic survey figure \ref{fig:schematic} 
where asymmetric ground states can play a role.
The upper panels show barriers and the lower panels lifetimes.
The left panels show results from reflection symmetric calculations
and the right panels from calculations where asymmetric shapes were allowed.
The results are calculated 
with the parameterization SV-min.}
\end{figure}

Figure \ref{fig:sf-barrier-sym-asym} compares the systematics of
barriers (upper panels) and lifetimes (lower panels) with and without
allowing for asymmetric shapes in the region of relevant isotopes.
One spots a slightly increased fission stability in case of allowed
asymmetry in the feature that the stable regions are somewhat
extended. However, the overall trends and the general impression of
the plots of systematics remains  the same. The effect of asymmetric
shapes appears rather at a quantitative level. The example
demonstrates how robust the analysis of global trends is.

\begin{acknowledgments}
  This work was supported by the BMBF under contract 06 ER 9063, by
  the Office of Nuclear Physics, U.S. Department of Energy under
  Contract Nos. DE-FG02-96ER40963 and DE-FC02-09ER41583, by the
  ExtreMe Matter Institute EMMI in the framework of the Helmholtz
  Alliance HA216/EMMI, by the Deutsche Forschungsgemeinschaft through
  contract SFB~634, by the Helmholtz Internation Center for FAIR
  within the framework of the LOEWE program launched by the state of
  Hesse and by the Helmholtz Association through the Nuclear
  Astrophysics Virtual Institute (VH-VI-417). We thank H. Feldmeier
  and F.-K. Thielemann for valuable discussions. 
\end{acknowledgments}


\bibliography{SHF,SHF-JE,SHF-PGR}

\end{document}